\documentclass[twocolumn,aps,amsfonts,amsmath,superscriptaddress,nofootinbib,showpacs,showkeys,byrevtex,floatfix,preprintnumbers,prd]{revtex4}
\usepackage{graphicx}
\usepackage{amssymb}
\usepackage{hyperref}
\usepackage{bm} 
\usepackage{subfigure} 
\usepackage{rotating} 
\usepackage{color}
\def\be#1\ee{\begin{equation}#1\end{equation}}
\def\bal#1\eal{\begin{align}#1\end{align}}
\def\bmu#1\emu{\begin{multline}#1\end{multline}}
\def\bga#1\ega{\begin{gather}#1\end{gather}}
\newcommand{\ba}{\begin{array}}
\newcommand{\ea}{\end{array}}
\newcommand{\n}{\notag}

\renewcommand{\d}{\partial}
\renewcommand{\bf}{\mathbf}
\renewcommand{\cal}{\mathcal}

\begin{document}

\title{Color Coulomb Potential in Yang-Mills Theory 
from Hamiltonian Flows}
\date{\today}

\author{Markus \surname{Leder}}
\email[corresponding author: ]{leder@tphys.physik.uni-tuebingen.de}
%\affiliation{Institut f\"ur Theoretische Physik, Universit\"at T\"ubingen, Auf der Morgenstelle 14, 72076 T\"ubingen, Germany}
\author{Hugo \surname{Reinhardt}}
\affiliation{Institut f\"ur Theoretische Physik, Universit\"at T\"ubingen, Auf der Morgenstelle 14, 72076 T\"ubingen, Germany}
\author{Axel \surname{Weber}}
\affiliation{Instituto de F\'isica y Matem\'aticas, Universidad Michoacana de San Nicol\'as de Hidalgo, Edificio C-3, Ciudad Universitaria, 58040 Morelia, Michoac\'an, Mexico}
\author{Jan M. \surname{Pawlowski}}
\affiliation{Institut f\"ur Theoretische Physik, Universit\"at Heidelberg, Philosophenweg 16, 69120 Heidelberg, Germany}
\affiliation{ExtreMe Matter Institute EMMI, GSI Helmholtzzentrum f\"ur Schwerionenforschung, \\
Planckstr.\ 1, 64291 Darmstadt, Germany}

\pacs{12.38.Aw, 05.10.Cc, 11.10.Ef, 11.15.Tk}
\keywords{functional renormalization group, color Coulomb potential, Yang-Mills theory}

\begin{abstract}
We consider the Hamiltonian formulation of Yang-Mills theory in the Coulomb
gauge and apply the recently developed technique of Hamiltonian flows. We
formulate a flow equation for the color Coulomb potential which allows for a scaling solution that results in an
almost linearly rising confining potential.
\end{abstract}

\maketitle

A profound understanding of the confinement mechanism in QCD still represents,
after almost 40 years of intense research, one of the most important
challenges in modern theoretical particle physics. We here report on a new
approach to the subject which uses the recently developed technique of
Hamiltonian flows \cite{LPR10}. The general setup is the Hamiltonian
formulation of Yang-Mills theory in the Coulomb gauge \cite{CL80}. Important
progress has been made over the last decade in this formulation, mainly via
the variational principle 
\cite{Sch85,Sch89,SS01,Zwa04,FR04a,FR04b,RF05,SLR06,ERS07,Rei08}.
The horizon condition is implemented in accord with the Gribov-Zwanziger
confinement scenario \cite{Gri78,Zwa91}, and scaling behavior of the 
equal-time two-point correlation functions results together with an infrared 
fixed point of an appropriately defined running coupling constant 
\cite{SLR06}.

The potential between static color sources in the large-distance or infrared 
limit is usually approximated by the color Coulomb potential.
The color Coulomb potential is given by the vacuum expectation value
\be
\left\langle (-\d D)^{-1} (-\d^2) (-\d D)^{-1} \right\rangle
\ee 
[see Eq.\ \eqref{covder} below for our definition of the covariant derivative 
$D$] and is usually expressed as
\be
\left\langle (-\d D)^{-1} \right\rangle (-\d^2) f (-\d^2)
\left\langle (-\d D)^{-1} \right\rangle \,,
\ee 
with the so-called Coulomb form factor $f$, which satisfies a 
Dyson-Schwinger equation (DSE). In order to calculate the color Coulomb
potential, in Ref.\ \cite{Zwa04} the Coulomb form
factor was simply set equal to one, while in
Refs.\ \cite{FR04a,FR04b,ERS07} the DSE for this form factor
was approximated by replacing in the loop integral the full ghost 
propagator $\left\langle (-\d D)^{-1} \right\rangle$ with the bare one, which 
results in an infrared finite Coulomb form factor. In this way, a strictly
linear growth of the color Coulomb potential with the distance between the 
color sources (for sufficiently large distances) has been found in Ref.\ 
\cite{ERS07}. It would now be natural to try to improve the approximation 
by using the full DSE for the Coulomb form factor. However, it turns out
\cite{FR04a,ERS08} that the full DSE for the Coulomb form factor cannot 
be consistently solved together with the DSEs for the static (equal-time)
gluon and ghost propagators with an infrared-divergent ghost form factor, 
i.e., implementing the horizon condition. In other words, a confining color
Coulomb potential cannot be obtained within the present approximation if the 
full DSE for the Coulomb form factor is used.

In the present letter, we will focus on the determination
of the color Coulomb potential with the help of a
different functional technique, the Hamiltonian flows \cite{LPR10}.
Interestingly, a consistent solution which exhibits scaling behavior 
of the static propagators and the color Coulomb potential is
readily found in this framework, without any additional approximation
for the Coulomb form factor.

The organization of this letter is as follows: we will start with a
brief presentation of the Hamiltonian flow technique and summarize the 
results of Ref.\ \cite{LPR10}. We then derive the flow equation for the
color Coulomb potential and, finally, present and discuss its solution.

The construction of the Hamiltonian flows starts from the generating
functional for Green's functions at equal times,
\be
Z[J] = \int \cal{D} A \, \text{det} (-\d D) |\psi [A]|^2 \,
e^{J \cdot A} \,,
\label{genfun}
\ee
where $D$ denotes the covariant derivative
\be
D^{ab}_i = \delta^{ab} \d_i - g f^{abc} A^c_i \label{covder}
\ee
and the dot in $J \cdot A$ stands for the contraction of color and spatial
indices and the integral over position or momentum,
\be
J \cdot A = \int d^3 x \, J^a_i (\bf{x}) A^a_i (\bf{x}) =
\int \frac{d^3 p}{(2\pi)^3} \, J^a_i (-\bf{p}) A^a_i (\bf{p}) \,.
\ee
In Eq.\ \eqref{genfun}, $\psi[A]$ represents the vacuum wave functional, $\text{det} (-\d D)$ is
the Faddeev-Popov determinant in Coulomb gauge \cite{CL80}, and the functional integral is 
over the transverse spatial gauge fields that fulfill the Coulomb gauge condition
$\d_i A^a_i = 0$ (we denote the contravariant spatial indices as
\emph{sub}indices).

We now write the Faddeev-Popov determinant in the well-known way as a
functional integral over ghost fields and introduce additional source fields 
for the latter. Furthermore, we introduce an infrared regulator term
$\Delta S_k$ with infrared cutoff scale $k$. Then the $k$-dependent 
generating functional becomes
\be
Z_k [J, \sigma, \bar{\sigma}] = \int \cal{D} A \cal{D} \bar{c} \cal{D} c \, 
e^{-S - \Delta S_k + J \cdot A + \bar{\sigma} \cdot c + \bar{c} \cdot \sigma}
\label{genfunk}
\ee
with the ``action''
\be
S = -\ln |\psi[A]|^2 + \int d^3 x \, \bar{c}^a (\bf{x}) (-\d D)^{ab}
c^b (\bf{x}) \,.
\ee
The regulator term is
\bal
\Delta S_k [A, c, \bar{c}] &= \frac{1}{2} \int \frac{d^3 p}{(2 \pi)^3} \,
A_i^a (-\bf{p}) R_{A,k} (p) A_i^a (\bf{p}) \n \\
&\phantom{=} {}+ \int \frac{d^3 p}{(2 \pi)^3} \, \bar{c}^a (-\bf{p}) g
\bar{R}_{c,k} (p) c^a (\bf{p}) \:,
\eal
with the regulator functions chosen in the present work as
\bal
R_{A,k} (p) &= 2 p \exp \left( \frac{k^2}{p^2} 
- \frac{p^2}{k^2} \right) \,, \n \\
\bar{R}_{c,k} (p) &= p^2 \exp \left( \frac{k^2}{p^2} 
- \frac{p^2}{k^2} \right) \,. \label{regfunct}
\eal
Here and in the following, we use the notation $p = |\bf{p}|$. The change
of $Z_k$ as defined in Eq.\ \eqref{genfunk} under a change of $k$
constitutes a (functional) renormalization group transformation.

In the limit of large $k$, the functional integration in Eq.\ \eqref{genfunk}
is restricted to ultraviolet modes where asymptotic freedom sets in, so that
$Z_k$ can be evaluated in perturbation theory. In the other extreme, for
$k \to 0$, the cutoff term $\Delta S_k$ vanishes and $Z_k \to Z$, the
full generating functional. Between these limits, the change of $Z_k$ with
$k$ obeys an exact integro-differential equation, the functional
renormalization group equation or flow equation.

In practice, it is more advantageous to consider the $k$-dependent effective
action $\Gamma_k$, the generating functional of one-particle irreducible 
Green's functions, instead of $Z_k$. We then approximate the corresponding 
exact flow equation for the variation of $\Gamma_k$ with $k$, by
approximating the form of $\Gamma_k$. Our basic assumption is that there
exists a solution which shows scaling behavior in the infrared.

We parameterize the static propagators as
\bal
\lefteqn{(2 \pi)^6 \left. \frac{\delta^2 \ln Z_k}
{\delta J_i^a (-\bf{p}) \delta J_j^b (\bf{q})} 
\right|_{J = \sigma = \bar{\sigma} = 0}} \hspace{2cm} \n \\
&= G_{A,k} (p) \delta^{ab} t_{ij} (\bf{p}) (2 \pi)^3 \delta (\bf{p} - \bf{q})
\,, \n \\
\lefteqn{-(2 \pi)^6 \left. \frac{\delta^2 \ln Z_k}
{\delta \bar{\sigma}^a (-\bf{p}) \delta \sigma^b (\bf{q})} 
\right|_{J = \sigma = \bar{\sigma} = 0}} \hspace{2cm} \n \\
&= \frac{1}{g} \, \bar{G}_{c,k} (p) \delta^{ab} 
(2 \pi)^3 \delta (\bf{p} - \bf{q}) \,, \label{proppar}
\eal
with the functions
\bal
G_{A,k} (p) &= \frac{1}{2 \omega_k (p) + R_{A,k} (p)} \;, \n \\
\bar{G}_{c,k} (p) &= \frac{1}{p^2/d_k (p) + \bar{R}_{c,k} (p)} \;.
\label{omddefs}
\eal
In the first of Eqs.\ \eqref{proppar}, $t_{ij} (\bf{p})$ denotes the
transverse projector or spatially transverse Kronecker delta. Our assumption
of scaling then refers to a power behavior for small momenta of the functions
$\omega_{k=0} (p)$ and $d_{k=0} (p)$,
\be
\omega_0 (p \to 0) \propto p^{-\alpha} \,, \qquad
d_0 (p \to 0) \propto p^{-\beta} \,. \label{powers}
\ee

The scaling assumption has a number of important consequences, just as its
counterpart in Landau gauge. As shown in Refs.\ 
\cite{Zwa04,FR04b,SLR06,FP07,FP09}, the exponents fulfill
\be
\alpha = 2 \beta - 1 \,, \qquad \alpha \ge - 3/2 \,. \label{sumrule}
\ee
In particular, in the infrared the diagrams with the greatest number of ghost 
loops dominate over the others (ghost dominance). Furthermore, the dressing
of the ghost-gluon vertex is constant \cite{FS05,Watson:2006yq}, 
usually referred to as the nonrenormalization of the vertex (see also
Ref.\ \cite{LS02}), so that we can calculate with a bare
ghost-gluon vertex, possibly adjusting the value of the coupling constant.
Finally, we can neglect vertices with more than two gluon lines, as far
as the dominant infrared behavior is concerned. 

\begin{figure}%[t] 
%$\ds \frac{\d}{\d k}$ \parbox{35pt}{\includegraphics[width=35pt,clip=]{gluon-propagator.eps}}$^{-1}=$
%$\;-\;$\parbox{63pt}{\includegraphics[width=63pt,clip=]{ghost-loop-1-trunc.eps}}
%$\;-\;$\parbox{63pt}{\includegraphics[width=63pt,clip=]{ghost-loop-2-trunc.eps}}
%{\Large$\ds \frac{\d}{\d k}$} \parbox{50pt}{\includegraphics[width=50pt,clip=]{gluon-propagator.eps}}{\Large$^{-1}=$}
%{\Large$\;-\;$}\parbox{90pt}{\includegraphics[width=90pt,clip=]{ghost-loop-1-trunc.eps}}
%{\Large$\;-\;$}\parbox{90pt}{\includegraphics[width=90pt,clip=]{ghost-loop-2-trunc.eps}}
\includegraphics[height=0.05\textheight]{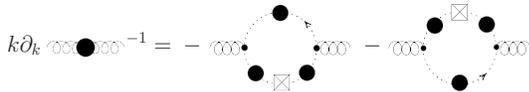}
\caption{Flow equation for the gluon propagator. The spiral and dotted lines 
  with the black circles denote the full gluon and ghost 
  propagators at cutoff momentum $k$, respectively. An insertion of the
  $k$-derivative $\d R_k/\d k$ of one of the regulator functions is represented
  by a square with a cross. In this and the following figure, the diagram
  on the left-hand side stands for the inverse propagator with
  the regulator function \emph{omitted}.} 
\label{gluonflow} 
\end{figure}

\begin{figure}%[t]  
%$\ds \frac{\d}{\d k}$ \parbox{35pt}{\includegraphics[width=35pt,clip=]{ghost-propagator.eps}}$^{-1}=$ 
%\parbox{49pt}{\includegraphics[width=49pt,clip=]{ghost-prop-loop-ghost-reg-trunc.eps}}
%$\;+\;$\parbox{49pt}{\includegraphics[width=49pt,clip=]{ghost-prop-loop-gluon-reg-trunc.eps}}
%{\Large$\ds \frac{\d}{\d k}$} \parbox{50pt}{\includegraphics[width=50pt,clip=]{ghost-propagator.eps}}{\Large$^{-1}=$} 
%\parbox{90pt}{\includegraphics[width=90pt,clip=]{ghost-prop-loop-ghost-reg-trunc.eps}}
%{\Large$\;+\;$}\parbox{90pt}{\includegraphics[width=90pt,clip=]{ghost-prop-loop-gluon-reg-trunc.eps}}
\includegraphics[height=0.05\textheight]{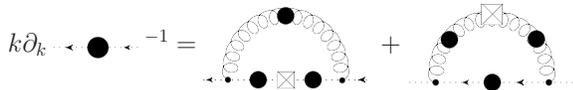}
\caption{Flow equation for the ghost propagator.} 
\label{ghostflow}
\end{figure}

Retaining only the contributions that are relevant to the infrared behavior,
the flow equation for $\Gamma_k$ induces the following integro-differential 
equations for the static propagators:
\begin{widetext} 
\bal
\frac{\d}{\d k} \, \omega_k (p) &= -\frac{N_c}{2} \int \frac{d^3 q}{(2 \pi)^3}
\bigg( \bar{G}_{c,k} \frac{\d \bar{R}_{c,k}}{\d k} \bar{G}_{c,k} \bigg) (q)
\bar{G}_{c,k} (|\bf{p} + \bf{q}|) q^2 
\left( 1 - (\hat{\bf{p}} \cdot \hat{\bf{q}})^2 \right) \,, \n \displaybreak[2] \\
\frac{\d}{\d k} \, d_k^{-1} (p) &= N_c \int \frac{d^3 q}{(2 \pi)^3}
\bigg[ \bigg( G_{A,k} \frac{\d R_{A,k}}{\d k} G_{A,k} \bigg) (q) 
\bar{G}_{c,k} (|\bf{p} + \bf{q}|) \n \\
&\phantom{= N_c \int \frac{d^3 q}{(2 \pi)^3} \Big[} {}+ 
\bigg( \bar{G}_{c,k} \frac{\d \bar{R}_{c,k}}{\d k} \bar{G}_{c,k} \bigg) (q)
G_{A,k} (|\bf{p} + \bf{q}|) \frac{q^2}{(\bf{p} + \bf{q})^2} \bigg]
\left( 1 - (\hat{\bf{p}} \cdot \hat{\bf{q}})^2 \right) \,, \label{propflows}
\eal
\end{widetext} 
where we have neglected in each equation the contribution of a tadpole term, 
in order to obtain a closed system of equations. The Eqs.\ \eqref{propflows}
are represented diagrammatically in Figs.\ \ref{gluonflow} and 
\ref{ghostflow}.
For details of the derivation of these equations, we refer the reader to Ref.\
\cite{LPR10}.

\begin{figure*}%[t] 
\hspace{\fill} 
\includegraphics[scale=0.35,clip=]{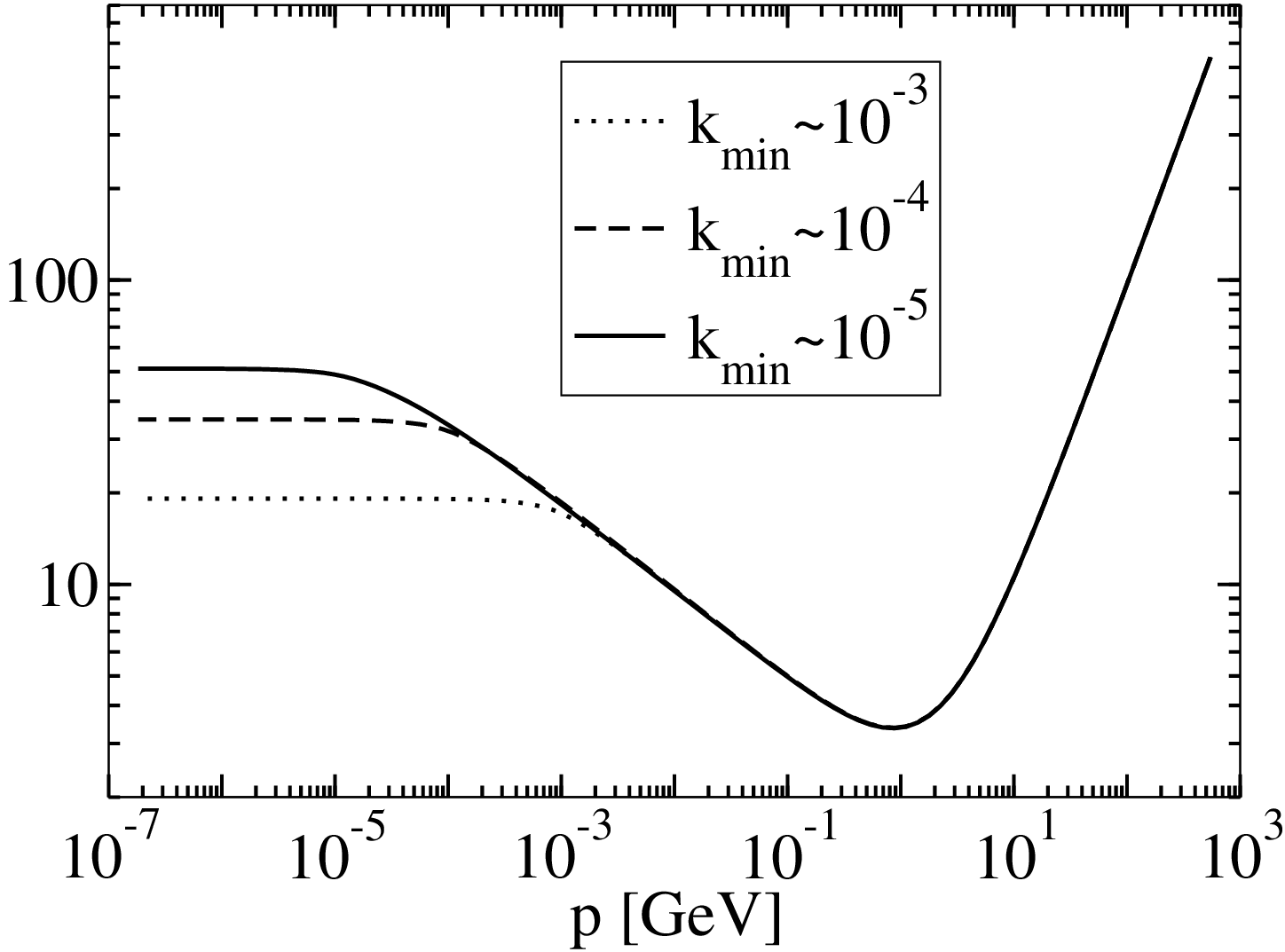} 
\hspace{\fill} 
\includegraphics[scale=0.35,clip=]{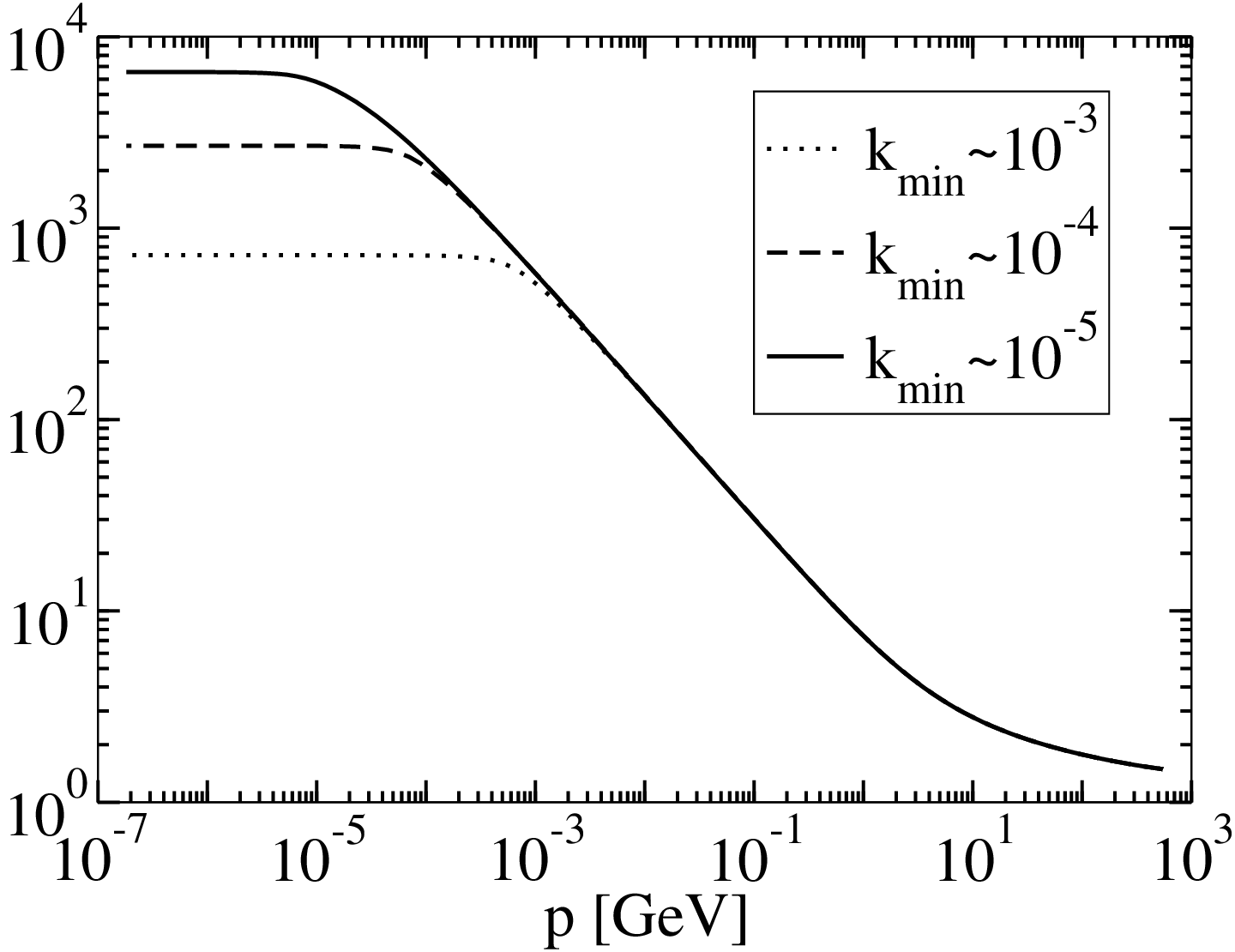} 
\hspace{\fill} 
\caption{Double-logarithmic plots of the inverse gluon propagator 
$\omega_k (p)$ (left) and the ghost dressing 
function $d_k (p)$ (right) resulting from a numerical solution of 
the flow equations \eqref{propflows},
for different minimal cutoffs $k = k_{\text{min}}$.
The scale on the momentum axes is an estimate obtained by matching
the position of the minimum of the inverse gluon propagator to the one
in Ref.\ \cite{BQR09}.
} 
\label{diffkmin} 
\end{figure*}

The solution of Eqs.\ \eqref{propflows} is determined by imposing initial
conditions on $\omega_k (p)$ and $d_k (p)$ 
at $k = \Lambda$, where $\Lambda$ is an arbitrary large momentum
scale. Here we adjust these initial conditions to satisfy the normalization 
conditions that $\omega_0 (p) \propto p$ for large momenta $p \lesssim 
\Lambda$, and that $d_0 (p)$ follow a power law in the infrared ($p \to 0$),
reflecting the properties of asymptotic freedom and infrared scaling,
respectively. It is possible to fulfill these normalization
conditions by adjusting the constants $a$ in $\omega_\Lambda (p) = a + p$
and $d_\Lambda (p) \equiv d_\Lambda$, technically upon rewriting
the differential equations 
\eqref{propflows} as integral equations over $k$. Again, we refer the reader
to Ref.\ \cite{LPR10} for details of the numerical procedure and here
only cite the result, represented in Fig.\ \ref{diffkmin}. 
Note that for technical
reasons, the equations can only be integrated down to a minimal value 
$k_{\text{min}}$ of $k$, and we show the dependence on the value of
$k_{\text{min}}$ in the figure.

In the infrared, for $p \gtrsim k_{\text{min}}$, we find the power behavior
of Eq.\ \eqref{powers}, with the numerical values for the exponents
\be
\alpha = 0.28 \,, \qquad \beta = 0.64 \,, \label{FRGsolution}
\ee
which conform to the sum rule \eqref{sumrule}, $\alpha = 2 \beta - 1$. Note
that a power behavior in the infrared (without specifying the exponent) was 
implemented as normalization condition for the ghost dressing function, but 
not for the gluon propagator. We also find an infrared fixed point for the
(running) effective fine structure constant,
\be
\alpha_s (p) = \frac{1}{4 \pi} \, \frac{p}{\omega_0 (p)} 
\left( d_0 (p) \right)^2 \to 1{.}70 \,. \label{alphasfp}
\ee

We can compare these results with the solutions of DSEs obtained in
the variational approach in Refs.\ \cite{FR04a,FR04b,ERS07}. In this
approach, two different solutions are found with $\beta_1 = 0{.}80$
\cite{FR04a,FR04b} and $\beta_2 = 1{.}00$ \cite{ERS07} and the
respective values of $\alpha$ according to the sum rule
\eqref{sumrule}. The same values of the exponents are also obtained
analytically in an approximation that is only adequate in the infrared
region \cite{Zwa04,SLR06}. It is not unexpected that the
functional renormalization group yields smaller values for the
critical exponents than the DSEs in corresponding approximations.  In
similar calculations in Landau gauge Yang-Mills theory it has been
found that the precise values of the exponents vary slightly with the
choice of the regulator functions. The upper limit of the interval
over which the values of the exponents vary, is given by the result of
the corresponding DSEs and is obtained from the functional
renormalization group for so-called ``optimized'' regulator functions
\cite{PLN04}.

In fact, we can simulate the use of optimized regulator functions by
simply replacing the functions $\omega_k (p)$ and $d_k (p)$ on the
right-hand sides of Eqs.\ \eqref{propflows} by $\omega_0 (p)$ and $d_0 (p)$.
We have argued in Ref.\ \cite{LPR10} that this replacement 
furthermore effectively takes into account part of the tadpole terms that 
we have omitted so far. With $\omega_0 (p)$ and $d_0 (p)$ on the
right-hand sides of Eqs.\ \eqref{propflows}, we can perform the integration 
over $k$ analytically and end up with equations very similar to those of the 
variational approach. We have shown in Ref.\ \cite{LPR10} that the
numerical solution of these equations matches almost perfectly the one of
Refs.\ \cite{FR04a,FR04b}, with $\beta = 0{.}80$. In this sense,
our solution \eqref{FRGsolution} corresponds to the solution 
$\beta_1 = 0{.}80$ of the variational (DS) approach.

We now come to the calculation of the color Coulomb potential in the
functional renormalization group approach. For its definition, one considers
the theory in the presence of an external static color charge density. Then
the color Coulomb potential is the vacuum expectation value of the part of the 
Hamiltonian that depends on the external color charges, explicitly in
momentum space (in an integral kernel notation)
\bal
F^{ab} (\bf{p}, -\bf{q}) &= \left\langle \langle \bf{p}, a |
(-\d D)^{-1} (-\d^2) (-\d D)^{-1} 
| \bf{q}, b \rangle \right\rangle \n \\
&= V_c (p) \delta^{ab} (2 \pi)^3 \delta (\bf{p} - \bf{q}) \,. \label{defcoul}
\eal
Introducing the composite operator
\be
K = \int \frac{d^3 k}{(2 \pi)^3} \, \bar{c}^d (-\bf{k}) k^2 c^d (\bf{k}) \,,
\label{K}
\ee
one can write
\be
F^{ab} (\bf{p}, -\bf{q}) = \left\langle c^a (\bf{p}) K \bar{c}^b (-\bf{q}) 
\right\rangle_{\text{GI}} \,. \label{Krepresent}
\ee
The label GI on the vacuum expectation value stands for gluon-irreducible,
meaning that one has to restrict the contributing diagrams to those where the
operator $K$ remains connected to the external points when all gluon lines
are cut.

The $k$-dependent color Coulomb potential $F_k^{ab} (\bf{p}, -\bf{q})$
is then naturally defined by including the cutoff term $\Delta S_k$ in
the functional integral representation of the vacuum expectation value
\eqref{Krepresent} as in Eq.\ \eqref{genfunk}, and a flow equation for 
$F_k$ can be derived in the standard way. For reasons of space, however,
here we present a much quicker and equivalent derivation of the
flow equation for $F_k$ which is based on the identity \cite{Swi88}
\be
\frac{\d}{\d g} \left[ g (-\d D)^{-1} \right] = 
(-\d D)^{-1} (-\d^2) (-\d D)^{-1} 
\ee
for the operators. We generalize this identity to
\bal
\lefteqn{\frac{\d}{\d g} \left[ g (-\d D + g \bar{R}_{c,k})^{-1} \right]}
\n \\ 
&= (-\d D + g \bar{R}_{c,k})^{-1} (-\d^2) (-\d D + g \bar{R}_{c,k})^{-1}
\label{genSwiftid} 
\eal
for our present purposes, so that
\bal
F_k^{ab} (\bf{p}, -\bf{q}) &= \big\langle \langle \bf{p}, a |
\frac{\d}{\d g} \left[ g (-\d D + g \bar{R}_{c, k})^{-1} \right]
| \bf{q}, b \rangle \big\rangle_k \n \\
&= V_{c, k} (p) \delta^{ab} (2 \pi)^3 \delta (\bf{p} - \bf{q}) \,.
\label{defVck}
\eal
Note that the rescaling of the ghost regulator
function with a factor of $g$ is essential to 
achieve a form equivalent to the definition of $F_k$ described above.

In order to put the identity \eqref{genSwiftid} to use inside the vacuum 
expectation values we are interested in, we define a $g$-derivative 
``at fixed integration measure'' 
\bal
\left. \d_g \right|_{\text{fm}} \langle \cal{O} \rangle_k
&= \int \cal{D} A \, \text{det} (-\d D + g \bar{R}_{c,k}) \left(
\frac{\d}{\d g} \cal{O} [A] \right) \n \\
&\phantom{=} {}\times |\psi [A]|^2 \exp \left( - \frac{1}{2} \, A \cdot
R_{A,k} \cdot A \right) \,,
\eal
for an arbitrary operator $\cal{O} [A]$. This definition immediately
implies that
\be
\left. \d_g \right|_{\text{fm}} G_{A, k} = 0 \,.
\ee
For the application of the
$g$-derivative to the static ghost propagator, we use the identity
\be
\left\langle c^a (\bf{p}) \bar{c}^b (-\bf{q}) \right\rangle_k
= \left\langle \langle \bf{p}, a | (- \d D + g \bar{R}_{c, k} )^{-1}
| \bf{q}, b \rangle \right\rangle_k \,.
\ee
With the help of definitions \eqref{proppar} and \eqref{defVck}, we 
then find
\be
\left. \d_g \right|_{\text{fm}} \bar{G}_{c, k} (p) = V_{c, k} (p) \,.
\ee
Introducing the \emph{Coulomb form factor} $f_k (p)$ by
\be
V_{c, k} (p) = \frac{1}{g^2} \bar{G}_{c, k} (p) p^2 f_k (p) \bar{G}_{c, k} (p) 
\label{deffk}
\,,
\ee
we may rewrite the latter identity as
\be
\left. \d_g \right|_{\text{fm}} d_k^{-1} (p) 
= \frac{1}{p^2} \left. \d_g \right|_{\text{fm}} \bar{G}_{c, k}^{-1} (p)
= - \frac{1}{g^2} f_k (p) \,.
\ee

As a consequence of these relations, we can derive a flow equation for the
Coulomb form factor by simply differentiating the flow equation 
\eqref{propflows} for $d_k^{-1}$ with respect to $g$ with the result
\begin{widetext}
\bal
\frac{\d}{\d k} \, f_k (p) &= -N_c \int \frac{d^3 q}{(2 \pi)^3}
\bigg[ \bigg( G_{A,k} \frac{\d R_{A,k}}{\d k} G_{A,k} \bigg) (q) \,
\bar{G}_{c,k}^2 (|\bf{p} + \bf{q}|) (\bf{p} + \bf{q})^2 
f_k (|\bf{p} + \bf{q}|) \n \\
&\phantom{= N_c \int \frac{d^3 q}{(2 \pi)^3} \Big[} {}+ 2 \bigg( 
\bar{G}_{c,k} \frac{\d \bar{R}_{c,k}}{\d k} \bar{G}^2_{c,k} \bigg) (q) \,
q^2 f_k (q) G_{A,k} (|\bf{p} + \bf{q}|) \frac{q^2}{(\bf{p} + \bf{q})^2} \bigg]
\left( 1 - (\hat{\bf{p}} \cdot \hat{\bf{q}})^2 \right) \,, \label{coulflow}
\eal
\end{widetext} 
where we have used the fact that $\left. \d_g \right|_{\text{fm}}$ and the
$k$-derivative commute. Since we have derived Eq.\ \eqref{coulflow} from 
Eq.\ \eqref{propflows}, several approximations are implicit in 
Eq.\ \eqref{coulflow} corresponding to those employed before 
in the derivation of the flow equation for $d_k^{-1}$. 

The standard derivation of the flow equation that makes use of the composite
operator $K$ [defined in Eq.\ \eqref{K}] and its equivalence with the 
argument presented above, as well
as an algebraic construction that avoids reference to gluon-irreducible
diagrams as in Eq.\ \eqref{Krepresent} will be detailed in a future 
publication.

Since the flow equations \eqref{propflows} for $\omega_k (p)$ and $d_k (p)$
do not involve $f_k (p)$, we can insert the solutions of the latter
equations found in Ref.\ \cite{LPR10} in Eq.\ \eqref{coulflow} and integrate
this flow equation applying the same techniques used before, i.e., we
convert Eq.\ \eqref{coulflow} into an integral equation and solve it 
numerically by an iterative procedure. Equation \eqref{coulflow} is linear 
and homogeneous in
$f_k (p)$, and we decided to normalize $f_k (p)$ to one in the ultraviolet 
(below the initial scale $\Lambda$) by appropriately adjusting the
initial condition $f_\Lambda (p) \equiv f_\Lambda$. It is clear from perturbation
theory that $f_k (p)$ should be constant in the ultraviolet (as long as 
$k \ll p$) except for logarithmic corrections.

\begin{figure*}%[t] 
\hspace{\fill} 
\includegraphics[scale=0.35,clip=]{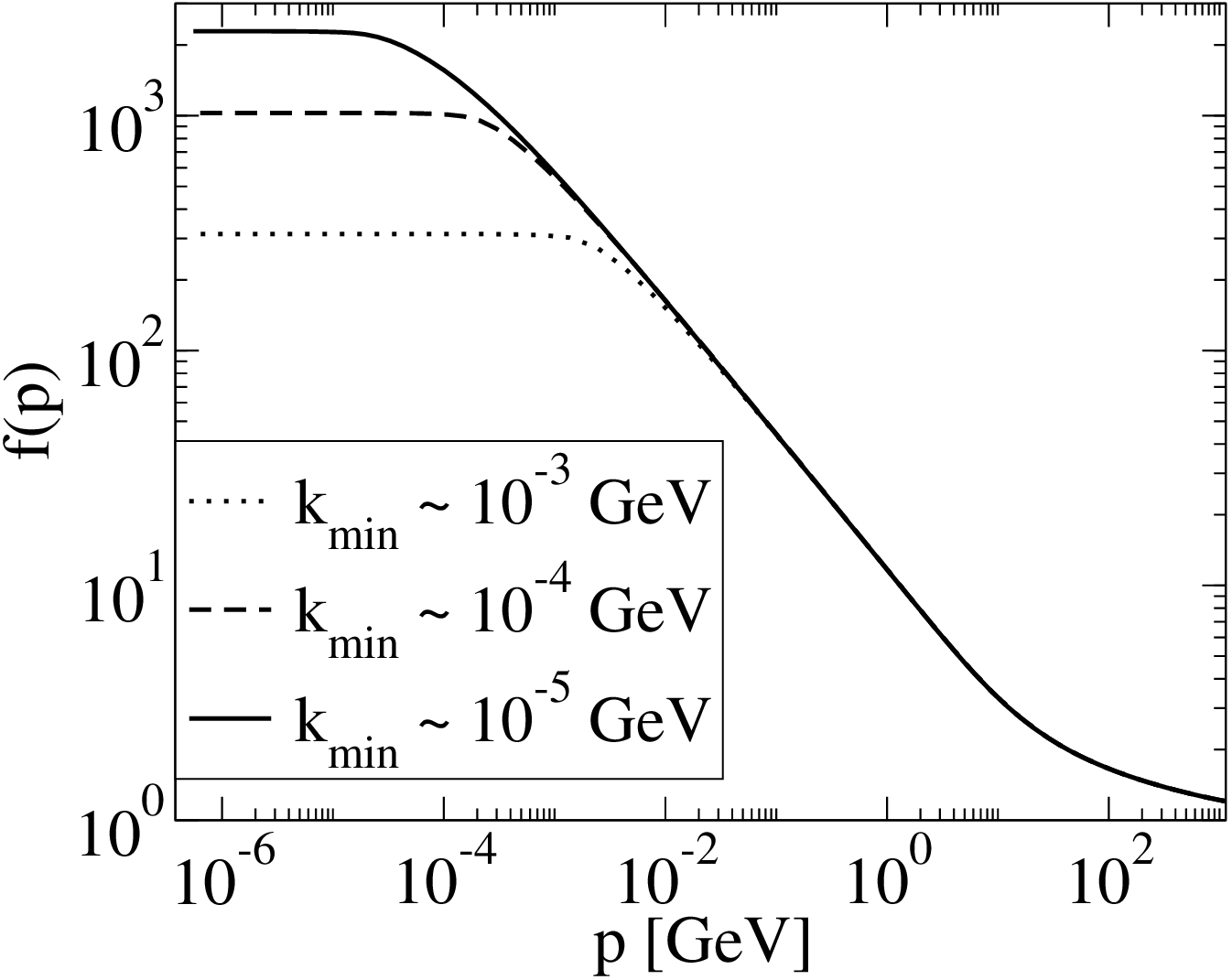} 
\hspace{\fill} 
\includegraphics[scale=0.35,clip=]{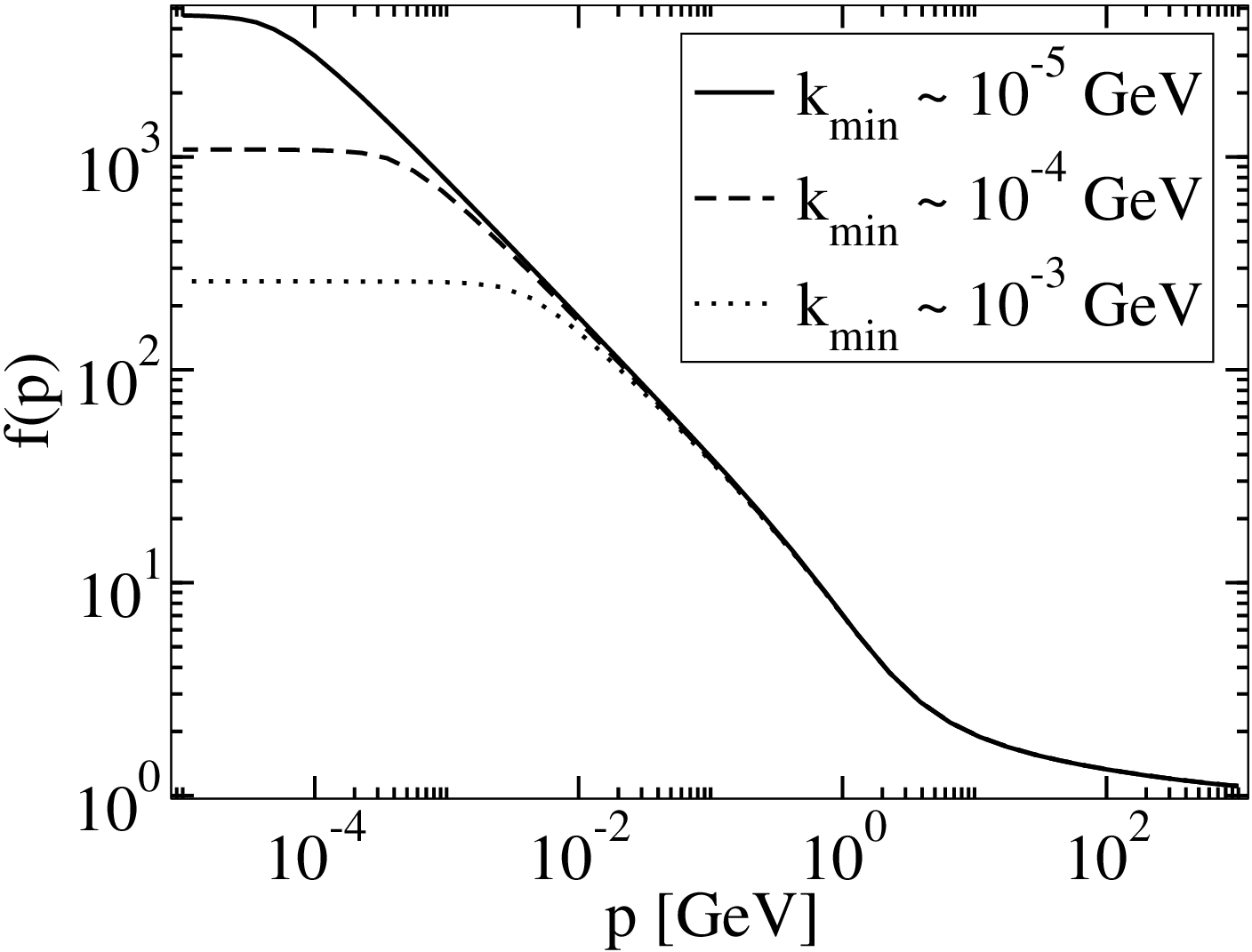} 
\hspace{\fill} 
%%%%%%%%%%%%%%%%%%%%%%%%%%%%%%%%%%%%%%
%\begin{figure*}%[t] 
%\begin{center}
%\includegraphics[scale=0.35,clip=]{coul_paper_GeV.eps} 
%\end{center}
%%%%%%%%%%%%%%%%%%%%%%%%%%%%%%%%%%%%%%%
\caption{The Coulomb form factor $f_k (p)$ as obtained from Eq.\ 
\eqref{coulflow} for different minimal cutoffs $k = k_{\text{min}}$,
calculated with the propagators from the flow equations \eqref{propflows}
(left) and the improved propagators (right).} 
\label{numresult} 
\end{figure*}

Somewhat surprisingly, and contrary to the negative result of the search
for a scaling solution in the variational approach complemented with 
DSEs \cite{ERS08}, a solution of Eq.\ \eqref{coulflow}
is readily found in the way described and is here represented in Fig.\
\ref{numresult}.
With the propagators obtained from the flow equations \eqref{propflows},
we get a power behavior
\be
f_k (p) \propto p^{-\gamma}
\ee 
in the infrared, with $\gamma = 0{.}57$ for $p \gtrsim k_{\text{min}}$. 
Consequently, the color Coulomb potential behaves as
\be
V_{c, k} (p) \propto p^{-\delta} \,, 
\qquad \delta = 2 + 2 \beta + \gamma
\ee
for $p \gtrsim k_{\text{min}}$, see Eqs.\ \eqref{omddefs} and \eqref{deffk}. 
Making use of our result \eqref{FRGsolution} for $\beta$,
we extrapolate $V_{c, k} (p)$ to
\be
V_c (p \to 0) \propto p^{-3{.}85} \label{eq:pot}
\ee
for $k=0$. 
We thus come quite close to a $(p^{-4})$-behavior which would correspond
to a potential that rises linearly with distance (for sufficiently large
distances). Also note that our present result $0{.}64$ for $\beta$ is
supposed to be smaller than the correct value as discussed below Eq.\
\eqref{alphasfp}, hence an improvement of the current approximation is 
expected to enhance the infrared exponent of $V_c (p)$.

The result \eqref{eq:pot} has been obtained with the propagators 
taken from the flow equations \eqref{propflows} that do not include
the tadpole diagrams. As argued before below Eq.\ \eqref{alphasfp},
we can easily take a part of the tadpole contributions into account in an 
effective way by replacing $\omega_k (p)$ and $d_k (p)$ on the right-hand 
sides of Eq.\ \eqref{propflows} with $\omega_0 (p)$ and $d_0 (p)$. In 
Ref.\ \cite{LPR10} we have also computed the flow of this improved 
truncation leading to $\beta = 0{.}80$. Inserting the latter solution 
into the flow equation \eqref{coulflow} for the Coulomb form factor, we 
are led to Fig.\ \ref{numresult} (right). The resulting infrared 
potential reads 
\be
V_c (p \to 0) \propto p^{-4{.}25} \,.
\label{eq:potim}
\ee
The exponents in Eqs.\ \eqref{eq:pot} and \eqref{eq:potim} provide us 
with an estimate for the systematic error of the present approximation:
\be
V_c (p \to 0) \propto p^{-\delta} \quad \text{with} 
\quad \delta \in [3{.}85\,,\, 4{.}25]\,, \label{eq:potfin}\ee
including $\delta =4$.

%%
%\begin{figure*}%[t] 
%\begin{center}
%\includegraphics[scale=0.35,clip=]{coul_ff_dse_input_back.eps} 
%\end{center}
%\caption{The Coulomb form factor $f_k (p)$ as obtained from Eq.\ 
%\eqref{coulflow} for different minimal cutoffs $k = k_{\text{min}}$, and the improved propagators.} 
%\label{numresultim} 
%\end{figure*} 
%%

In summary, the method of Hamiltonian flows allows for scaling
solutions for the static two-point functions and the color Coulomb
potential without additional approximations for the Coulomb form
factor, contrary to variational approaches.  In particular, we find an
infrared-divergent Coulomb form factor and an almost linearly confining 
potential $V_c (p) \propto p^{-\delta}$ with $\delta \in
[3{.}85\,,\, 4{.}25]$.
We expect that an improvement of the approximation employed narrows the 
above interval for $\delta$ while still including $\delta = 4$.

\begin{acknowledgments}
M.L. was supported by the Internationales Graduiertenkolleg 
``Hadronen im Vakuum, in Kernen und Sternen''. H.R. acknowledges support by 
DFG-Re856/6-3. A.W. is grateful to CIC-UMSNH for financial support.
J.M.P. acknowledges support by Helmholtz Alliance HA216/EMMI.
\end{acknowledgments}

\end{document}